\documentclass[preprint,3p]{elsarticle}
\usepackage{amssymb}
\usepackage{epsfig}

\journal{Astroparticle Physics}

\begin{document}

\begin{frontmatter}

\title{
Zenith distribution and flux of atmospheric muons measured with the 5-line ANTARES detector
}

\def\elsauthors{ANTARES Collaboration \\ \vspace{0.4cm}}

\author[IFIC]{J.A. Aguilar}
\author[Colmar]{A. Albert}
\author[Erlangen]{G. Anton}
\author[IRFU/SEDI]{S. Anvar}
\author[UPV]{M. Ardid}
\author[NIKHEF]{A.C. Assis Jesus}
\author[NIKHEF]{T.~Astraatmadja\fnref{tag:1}}
\author[CPPM]{J-J. Aubert}
\author[Erlangen]{R. Auer}
\author[APC]{B. Baret}
% ...

\fntext[tag:1]{\scriptsize{Also at University of Leiden, the Netherlands}}
\fntext[tag:2]{\scriptsize{On leave at DESY, Platanenallee 6, D-15738 Zeuthen, Germany}}

\cortext[cor1]{Corresponding author}

\newpage
\nopagebreak[3]
\address[IFIC]{\scriptsize{IFIC - Instituto de F\'isica Corpuscular, Edificios Investigaci\'on de Paterna, CSIC - Universitat de Val\`encia, Apdo. de Correos 22085, 46071 Valencia, Spain}}\vspace*{0.15cm}
\nopagebreak[3]
\vspace*{-0.20\baselineskip}
\nopagebreak[3]
\address[Colmar]{\scriptsize{GRPHE - Institut universitaire de technologie de Colmar, 34 rue du Grillenbreit BP 50568 - 68008 Colmar, France }}\vspace*{0.15cm}
\nopagebreak[3]
\vspace*{-0.20\baselineskip}
\nopagebreak[3]
\address[Erlangen]{\scriptsize{Friedrich-Alexander-Universit\"{a}t Erlangen-N\"{u}rnberg, Erlangen Centre for Astroparticle Physics, Erwin-Rommel-Str. 1, 91058 Erlangen, Germany}}\vspace*{0.15cm}
\nopagebreak[3]
\vspace*{-0.20\baselineskip}
\nopagebreak[3]
\address[IRFU/SEDI]{\scriptsize{Direction des Sciences de la Mati\`ere - Institut de recherche sur les lois fondamentales de l'Univers - Service d'Electronique des D\'etecteurs et d'Informatique, CEA Saclay, 91191 Gif-sur-Yvette Cedex, France}}\vspace*{0.15cm}
\nopagebreak[3]
\vspace*{-0.20\baselineskip}
\nopagebreak[3]
\address[UPV]{\scriptsize{Institut d'Investigaci\'o per a la Gesti\'o Integrada de Zones Costaneres (IGIC) - Universitat Polit\`ecnica de Val\`encia. C/ Paranimf, 1. E-46730 Gandia, Spain.}}\vspace*{0.15cm}
\nopagebreak[3]
\vspace*{-0.20\baselineskip}
\nopagebreak[3]
\address[NIKHEF]{\scriptsize{FOM Instituut voor Subatomaire Fysica Nikhef, Science Park 105, 1098 XG Amsterdam, The Netherlands}}\vspace*{0.15cm}
\nopagebreak[3]
\vspace*{-0.20\baselineskip}
\nopagebreak[3]
\address[CPPM]{\scriptsize{CPPM - Centre de Physique des Particules de Marseille, CNRS/IN2P3 et Universit\'e de la M\'editerran\'ee, 163 Avenue de Luminy, Case 902, 13288 Marseille Cedex 9, France}}\vspace*{0.15cm}
\nopagebreak[3]
\vspace*{-0.20\baselineskip}
\nopagebreak[3]
\address[APC]{\scriptsize{APC - Laboratoire AstroParticule et Cosmologie, UMR 7164 (CNRS, Universit\'e Paris 7 Diderot, CEA, Observatoire de Paris) 10, rue Alice Domon et L\'eonie Duquet 75205 Paris Cedex 13,  France}}\vspace*{0.15cm}
\nopagebreak[3]
\vspace*{-0.20\baselineskip}
\nopagebreak[3]
\address[LAM]{\scriptsize{LAM - Laboratoire d ' Astrophysique de Marseille, P\^ole de l ' \'Etoile Site de Ch?teau-Gombert, rue Fr\'ed\'eric Joliot-Curie 38,  13388 Marseille cedex 13, France }}\vspace*{0.15cm}
\nopagebreak[3]
\vspace*{-0.20\baselineskip}
\nopagebreak[3]
\address[Bologna-UNI]{\scriptsize{Dipartimento di Fisica dell'Universit\`a, Viale Berti Pichat 6/2, 40127 Bologna, Italy}}\vspace*{0.15cm}
\nopagebreak[3]
\vspace*{-0.20\baselineskip}
\nopagebreak[3]
\address[Bologna]{\scriptsize{INFN - Sezione di Bologna, Viale Berti Pichat 6/2, 40127 Bologna, Italy}}\vspace*{0.15cm}
\nopagebreak[3]
\vspace*{-0.20\baselineskip}
\nopagebreak[3]
\address[Roma-UNI]{\scriptsize{Dipartimento di Fisica dell'Universit\`a La Sapienza, P.le Aldo Moro 2, 00185 Roma, Italy}}\vspace*{0.15cm}
\nopagebreak[3]
\vspace*{-0.20\baselineskip}
\nopagebreak[3]
\address[Rome]{\scriptsize{INFN -Sezione di Roma, P.le Aldo Moro 2, 00185 Roma, Italy}}\vspace*{0.15cm}
\nopagebreak[3]
\vspace*{-0.20\baselineskip}
\nopagebreak[3]
\address[Pisa-UNI]{\scriptsize{Dipartimento di Fisica dell'Universit\`a, Largo B. Pontecorvo 3, 56127 Pisa, Italy}}\vspace*{0.15cm}
\nopagebreak[3]
\vspace*{-0.20\baselineskip}
\nopagebreak[3]
\address[Pisa]{\scriptsize{INFN - Sezione di Pisa, Largo B. Pontecorvo 3, 56127 Pisa, Italy}}\vspace*{0.15cm}
\nopagebreak[3]
\vspace*{-0.20\baselineskip}
\nopagebreak[3]
\address[INAF]{\scriptsize{INAF-IASF, via P. Gobetti 101, 40129 Bologna, Italy}}\vspace*{0.15cm}
\nopagebreak[3]
\vspace*{-0.20\baselineskip}
\nopagebreak[3]
\address[GEOAZUR]{\scriptsize{G\'eoazur - Universit\'e de Nice Sophia-Antipolis, CNRS/INSU, IRD, Observatoire de la C\^ote d'Azur and Universit\'e Pierre et Marie Curie, F-06235, BP 48, Villefranche-sur-mer, France}}\vspace*{0.15cm}
\nopagebreak[3]
\vspace*{-0.20\baselineskip}
\nopagebreak[3]
\address[Bari]{\scriptsize{INFN - Sezione di Bari, Via E. Orabona 4, 70126 Bari, Italy}}\vspace*{0.15cm}
\nopagebreak[3]
\vspace*{-0.20\baselineskip}
\nopagebreak[3]
\address[LNS]{\scriptsize{INFN - Laboratori Nazionali del Sud (LNS), Via S. Sofia 62, 95123 Catania, Italy}}\vspace*{0.15cm}
\nopagebreak[3]
\vspace*{-0.20\baselineskip}
\nopagebreak[3]
\address[Genova]{\scriptsize{INFN - Sezione di Genova, Via Dodecaneso 33, 16146 Genova, Italy}}\vspace*{0.15cm}
\nopagebreak[3]
\vspace*{-0.20\baselineskip}
\nopagebreak[3]
\address[IRFU/SPP]{\scriptsize{Direction des Sciences de la Mati\`ere - Institut de recherche sur les lois fondamentales de l'Univers - Service de Physique des Particules, CEA Saclay, 91191 Gif-sur-Yvette Cedex, France}}\vspace*{0.15cm}
\nopagebreak[3]
\vspace*{-0.20\baselineskip}
\nopagebreak[3]
\address[COM]{\scriptsize{COM - Centre d'Oc\'eanologie de Marseille, CNRS/INSU et Universit\'e de la M\'editerran\'ee, 163 Avenue de Luminy, Case 901, 13288 Marseille Cedex 9, France}}\vspace*{0.15cm}
\nopagebreak[3]
\vspace*{-0.20\baselineskip}
\nopagebreak[3]
\address[UPS]{\scriptsize{Universit\'e Paris-Sud 11 - D\'epartement de Physique - F - 91403 Orsay Cedex, France}}\vspace*{0.15cm}
\nopagebreak[3]
\vspace*{-0.20\baselineskip}
\nopagebreak[3]
\address[Genova-UNI]{\scriptsize{Dipartimento di Fisica dell'Universit\`a, Via Dodecaneso 33, 16146 Genova, Italy}}\vspace*{0.15cm}
\nopagebreak[3]
\vspace*{-0.20\baselineskip}
\nopagebreak[3]
\address[IPHC]{\scriptsize{IPHC-Institut Pluridisciplinaire Hubert Curien - Universit\'e de Strasbourg et CNRS/IN2P3   23 rue du Loess -BP 28-  F67037 Strasbourg Cedex 2}}\vspace*{0.15cm}
\nopagebreak[3]
\vspace*{-0.20\baselineskip}
\nopagebreak[3]
\address[NIOZ]{\scriptsize{Royal Netherlands Institute for Sea Research (NIOZ), Landsdiep 4,1797 SZ 't Horntje (Texel), The Netherlands}}\vspace*{0.15cm}
\nopagebreak[3]
\vspace*{-0.20\baselineskip}
\nopagebreak[3]
\address[KVI]{\scriptsize{Kernfysisch Versneller Instituut (KVI), University of Groningen, Zernikelaan 25, 9747 AA Groningen, The Netherlands}}\vspace*{0.15cm}
\nopagebreak[3]
\vspace*{-0.20\baselineskip}
\nopagebreak[3]
\address[UU]{\scriptsize{Universiteit Utrecht, Faculteit Betawetenschappen, Princetonplein 5, 3584 CC Utrecht, The Netherlands}}\vspace*{0.15cm}
\nopagebreak[3]
\vspace*{-0.20\baselineskip}
\nopagebreak[3]
\address[UvA]{\scriptsize{Universiteit van Amsterdam, Instituut voor Hoge-Energie Fysika, Science Park 105, 1098 XG Amsterdam, The Netherlands}}\vspace*{0.15cm}
\nopagebreak[3]
\vspace*{-0.20\baselineskip}
\nopagebreak[3]
\address[Catania-UNI]{\scriptsize{Dipartimento di Fisica ed Astronomia dell'Universit\`a, Viale Andrea Doria 6, 95125 Catania, Italy}}\vspace*{0.15cm}
\nopagebreak[3]
\vspace*{-0.20\baselineskip}
\nopagebreak[3]
\address[WIN]{\scriptsize{University of Wisconsin - Madison, 53715, WI, USA}}\vspace*{0.15cm}
\nopagebreak[3]
\vspace*{-0.20\baselineskip}
\nopagebreak[3]
\address[ISS]{\scriptsize{Institute for Space Sciences, R-77125 Bucharest, M\u{a}gurele, Romania     }}\vspace*{0.15cm}
\nopagebreak[3]
\vspace*{-0.20\baselineskip}
\nopagebreak[3]
\address[ITEP]{\scriptsize{ITEP - Institute for Theoretical and Experimental Physics, B. Cheremushkinskaya 25, 117218 Moscow, Russia}}\vspace*{0.15cm}
\nopagebreak[3]
\vspace*{-0.20\baselineskip}
\begin{abstract}
The ANTARES high energy neutrino telescope is a three-dimensional array 
of about 900 photomultipliers distributed 
over 12 mooring lines installed in the Mediterranean Sea.  
Between February and November 2007 it acquired data in a 5-line configuration. 
The zenith angular distribution of the atmospheric muon flux 
and the associated depth-intensity relation are measured and compared 
with previous measurements and Monte Carlo expectations. 
An evaluation of the systematic effects due to uncertainties on 
environmental and detector parameters is presented.
\end{abstract}
\end{frontmatter}
\section{Introduction}
The ANTARES collaboration has recently completed the 
construction of a neutrino telescope in the Northern Hemisphere. 
Its location is about 40 km off the coast of Toulon, France, at a depth of 2475 m in the 
Mediterranean Sea (42$^\circ$48'N, 6$^\circ$10'E). 
The main goal of the experiment is the search for high-energy neutrinos from astrophysical sources 
such as Active Galactic Nuclei, microquasars, supernova remnants and gamma-ray bursters. 
A neutrino telescope in the Northern hemisphere includes the Galactic Centre 
in its field of view and is complementary to the 
AMANDA and IceCube Antarctic telescopes \cite{amanda}. 
Neutrinos are observed indirectly via the detection of the Cherenkov light 
emitted along the trajectory of high energy charged particles emerging from neutrino interactions. 
The Cherenkov light is collected by light sensors (photomultipliers - PMTs) distributed in a three-dimensional array.
The PMTs are arranged on twelve vertical lines. 
The first detection line was deployed in February 2006 and a second became operational in October of the same year. 
Three more lines were connected in January 2007 and five more in December 2007. 
The apparatus reached its complete configuration when the last two lines were deployed and connected in May 2008.\\

The telescope is optimised for the detection of muon neutrinos, 
since muons resulting from  charged current interactions can travel kilometres and 
are almost collinear with the parent neutrinos  at energies above 1 TeV. 
The technique relies on discriminating upward-going neutrino-induced muons from the 
flux of downward-going atmospheric muons, which represent the majority of events in the ANTARES detector. 
Atmospheric muons are produced mainly by the decay of charged pions and kaons resulting from the 
interaction of high energy cosmic rays with atomic nuclei in the  atmosphere. 
Although they are a background for neutrino detection, 
the atmospheric muons are useful to verify the detector response.\\

This paper reports on the determination of the zenith angular distribution and on the measurement of the 
flux of atmospheric muons using data taken between February and November 2007 when the detector comprised 5 lines. 
The experimental results are compared with atmospheric muon flux simulations, with previous ANTARES measurements obtained 
with the first detector line \cite{line1} and with the results of a low energy event selection \cite{zaborov}.

\section{Experimental setup and data selection}

The full ANTARES detector consists of 12 flexible lines, each with a total height of 450 m, 
separated by distances of 60-70 m from each other. 
They are anchored to the sea bed and kept near vertical by buoys at the top of the line. 
Each line carries a total of 75 10" Hamamatsu PMTs housed in glass spheres, 
the optical modules (OM)  \cite{ant02}, arranged in 25 storeys (3 optical modules per storey) separated by 14.5 m, 
starting 100 m above the sea floor. Each PMT is oriented 45$^\circ$ downward with respect to the vertical. 
A titanium cylinder in each storey houses the electronics for readout and control, 
together with compasses and tiltmeters used to measure the heading and the inclination of the storeys. 
To reconstruct the neutrino direction with high precision 
a good knowledge of PMT position and of hit arrival times is necessary. 
The positions of the optical modules are  measured by a system of acoustic transponders 
and receivers distributed over the lines and on the sea bed.  
A system of LED beacons housed in some of the storeys and a laser beacon located at the 
bottom of two of the lines are used for timing calibration \cite{ant07b}. 
The PMT signals are digitized by a custom built ASIC chip.
For analog pulses which are larger than a preset threshold, typically 1/3 photoelectron (p.e.), 
the arrival time  and the integrated charge of the pulse are measured (referred to as L0 hit) 
and the corresponding data are sent to shore. 
The data stream is processed by a computer farm in the shore station which searches 
for different physics signals according to predefined trigger conditions. 
The DAQ system is described in detail in Ref.  \cite{ant07c}. \\

During the 5-line data taking period, 
the trigger algorithm required at least 5 causally connected L1 hits, 
where an L1 hit is defined either as 2 L0 hits in coincidence within 20 ns in two 
optical modules on the same storey or as a single L0 hit with an amplitude larger than 3 p.e.\\ 

\begin{figure}
\begin{center}
\mbox{\includegraphics[width=15cm]{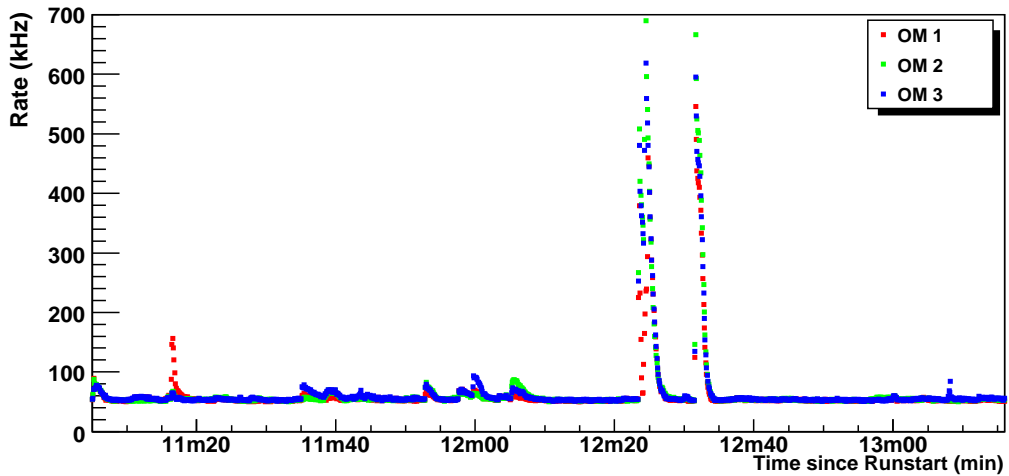}}
\caption{\label{countrate}\footnotesize {Typical counting rates in three optical modules of a detector line storey. }}
\end{center}
\end{figure}

In Figure \ref{countrate} the typical counting rate registered by an optical module is shown. 
It is characterized by a minimum constant rate, the so called baseline rate, 
originating from $^{40}$K decay and bioluminescent bacteria. 
Superimposed on the baseline are occasional bursts, of a few second duration, 
attributed to luminous emission by macro-organisms. \\

For this analysis, only runs fulfilling the following quality criteria are selected:
i) at least 300 out of 375 installed optical modules are active during a run (80\%); 
ii) the baseline rate is below 120 kHz; 
iii) the burst fraction is less than 20\%. 
The burst fraction is defined as the fraction of time during which the 
instantaneous optical background rate exceeds the baseline rate by at least 20\%.
With this selection the active detector time was about 90 effective days, 
during which more than $10^7$ atmospheric muon triggers were collected.

\section{Analysis chain}
The analysis chain comprises the simulation of the underwater muon flux induced 
by the interaction of the primary cosmic rays in the atmosphere, 
the simulation of the Cherenkov light in the instrumented volume and the 
reconstruction of the muon tracks starting from the detected or simulated signals.

\subsection{Simulation of the underwater muon flux}
Two different methods are used to describe the atmospheric muon flux reaching the detector. \\

The first approach is a Monte Carlo simulation based on CORSIKA version 6.2 \cite{corsika}. Two different models have been considered  for the description of hadronic interactions:
QGSJET.01 \cite{qgsjet} and SIBYLL \cite{sibyll}. 
Showers were simulated over an energy range from 1 to $10^5$ TeV/nucleon and zenith angles between $0^\circ$ and $85^\circ$.
For zenith angles larger than $70^\circ$ the curvature of the atmosphere is taken into account. 
Five mass groups (p, He, CNO, Mg-Si and Fe) are generated with an energy spectrum proportional to $E^{-2}$. More than $10^{10}$ air showers induced by primary cosmic ray nuclei were simulated with QGSJET.01. This corresponds to a detector live time ranging from few days to few years 
depending on the primary mass, its angle and energy. A smaller sample was simulated with SIBYLL, mainly to evaluate the effect of a different choice of the hadronic interaction model.
Muons reaching the sea level with energies larger than  500 GeV are propagated through 
sea water to the detector with the MUSIC package \cite{music}. 
The Monte Carlo events are weighted according to two different primary cosmic ray 
spectra : the H\"orandel/poly-gonato model (with rigidity dependent knee energies) \cite{horandel}   and the NSU model \cite{nsu,bugaev}. Both are based on experimental results, combining direct measurements and results from extensive air shower arrays. Note that in the energy range relevant for us (10 - few hundred TeV) the two all-particle spectra differ by about 30\%.

The second approach uses a Monte Carlo event generator based on a parametrized description of the underwater muon flux (MUPAGE) \cite{mupage}. 
The flux, the angular distribution, the multiplicity and the energy spectrum of muons are obtained 
from parametric formulae \cite{mupageY}.
A number of showers corresponding to one month of live time has been generated. \\

\subsection{Light production and detector simulation}
All muons reaching the detector depth with an energy above 20 GeV are transported  
through the detector active volume using MUSIC. 
As the  number of photons emitted along the muon trajectory 
is  large, the simulation of the light production is very time consuming. 
For this reason, a set of tables storing amplitude and time of 
Cherenkov photons hitting the PMTs is created.
They are constructed taking into account the water properties at the ANTARES site (light absorption length and scattering model), 
the characteristics of the optical modules (geometry and efficiency) and the position, 
distance and orientation of an optical module with respect to a given muon track or  electromagnetic shower.
The optical background is taken from the actual counting rate observed in the data. 
Dead channels and the efficiencies of the front-end electronics are also taken into account.

\subsection{Track Reconstruction}
A fit algorithm based on a chisquare-like minimization is used 
to reconstruct the tracks. 
It approximates each storey as a space point on a vertical detector line. 
The hits in the event are time ordered and merged on each storey if closer than 20 ns; 
in this case the time of the first hit on the storey is taken and the charges summed.  
In order to augment the weight of coincidences with respect to single high-charge pulses,
the summed charge is further increased by 1 if the hits originate 
from different optical modules of the same storey.  
All hits, merged or single, having a minimal charge of 2.5 p.e. are defined as step-1 hits. 
Next, a search is made for clusters of step-1 hits by requiring the presence of 
two step-1 hits within 80 ns on two adjacent floors or 
within 160 ns on two next-to-adjacent floors.  
For lines having one or more such clusters, single L0 hits within the line 
are selected if they are causally connected to the cluster(s) of the line. 
The times and the positions of all  selected hits on all lines
are then used in a  prefit of the track. The result 
of this prefit provides the starting values for the final chisquare minimization. 
This final fit yields two angles defining the track direction 
and the three coordinates of the track at a chosen time.\\

Monte Carlo studies indicate that  this reconstruction strategy 
yields a resolution on the zenith angle of $0.7^\circ$ for atmospheric muons.
The assumption of a vertical detector line introduces a further uncertainty on the reconstructed zenith angle. The sea currents registered during the selected data period induce displacements of the OMs from the nominal position which are smaller than one metre over 350 m height. This corresponds to an additional error of about $0.15^\circ$.\\

\begin{figure}
\begin{center}
\mbox{\includegraphics[width=13.7 cm]{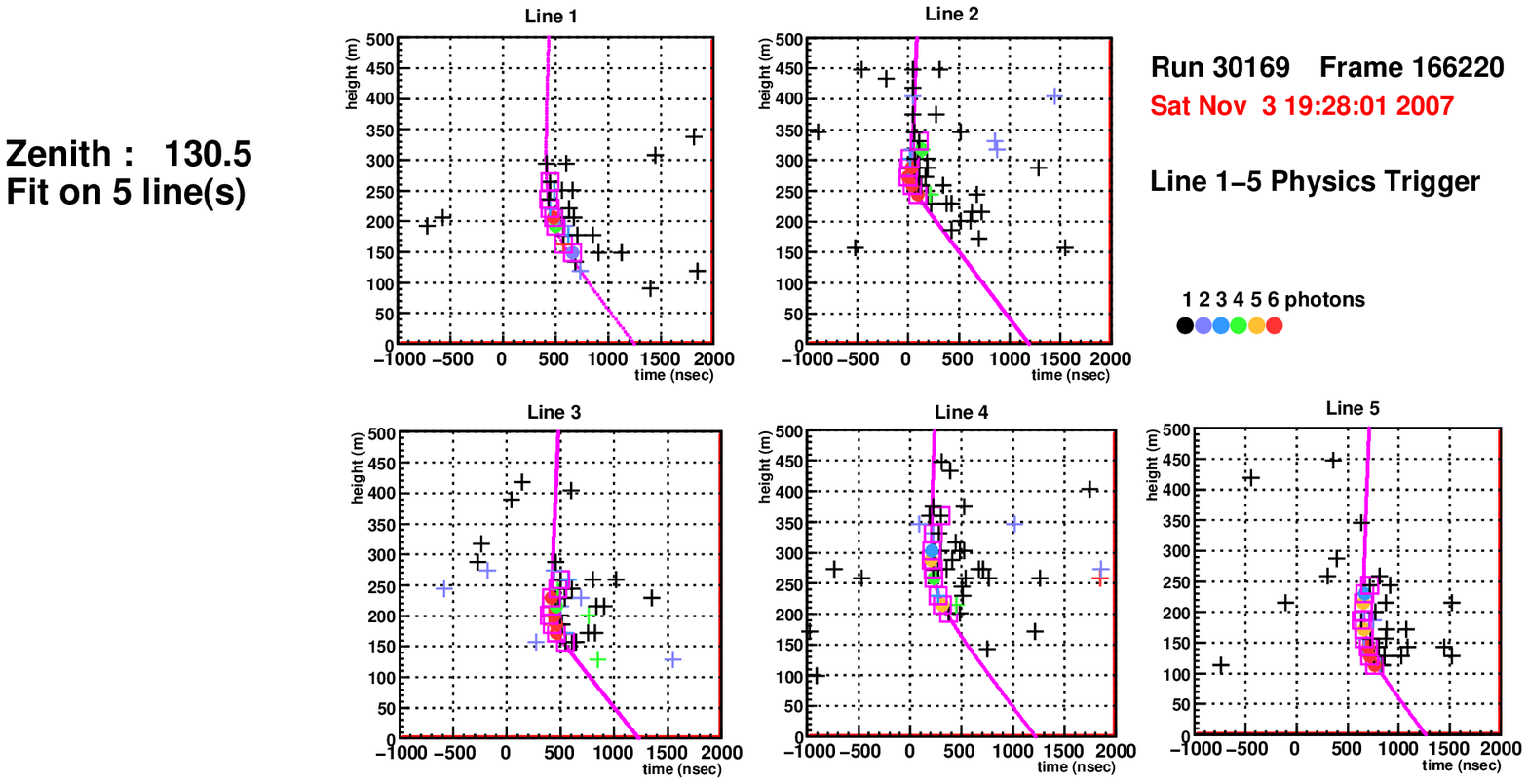}}
\caption{\label{mu_5lin}\footnotesize {A downgoing track, reconstructed using all 5 lines.  
In each display, the curve shows the reconstructed muon track, 
i.e. the z position of the Cherenkov cone, at the x-y position of the corresponding line, as a function of time.
Horizontal axis: time in nanoseconds; vertical axis: height above the sea-bed in metres. 
Crosses are hits in a time window of 3 microseconds around the trigger; 
full circles are hits passing the trigger condition (see Section 2); open boxes are hits used in the final fit. }}
\end{center}
\end{figure}

An example of a downward going track, reconstructed in the data, is shown in Figure \ref{mu_5lin}. 
The horizontal axis indicates hit time, while  the vertical axis shows  the height of the fired storey in metres. 
On this picture, crosses represent hits in a time window of 3 microseconds around the trigger, 
full circles indicate hits participating in the trigger and open boxes designate those hits which have been used in the fit. 
\begin{figure}
\begin{center}
\mbox{\includegraphics[width=11.5cm]{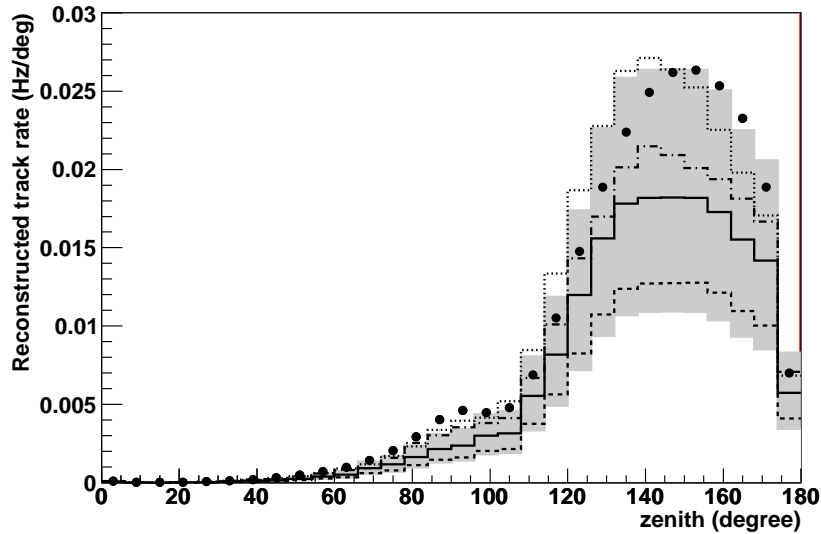}}
\caption{\label{zenith}\footnotesize {Zenith angle distribution of reconstructed tracks. 
Black points are the data.  The dotted line is the  MUPAGE Monte Carlo expectation.
The solid line  corresponds to the simulation with CORSIKA + QGSJET and  
the NSU model of the primary cosmic ray flux;  
the dashed  line is CORSIKA + QGSJET with the ``poly-gonato" model. 
The dashed-dotted line refers to CORSIKA + SIBYLL and  
the NSU model.
The shadowed band represents the systematic uncertainty, 
with respect to the solid line. See text for details.}}
\end{center}
\end{figure}

\section{Zenith distribution }

Figure \ref{zenith} shows the zenith angle distribution of the reconstructed tracks. Each track represents a muon bundle, regardless of the real number of muons in the bundle. Only tracks reconstructed using hits on at least 6 floors
are considered in this plot.
The systematic errors on the Monte Carlo expectations are due to the uncertainties 
on the description of the detector, on the knowledge of the environmental parameters 
and on the primary cosmic ray composition and hadronic interaction models. 
The effect of the first two terms is shown as a grey band in Figure \ref{zenith} 
and has been evaluated by repeating the Monte Carlo simulation 
described in Section 3.2 for various assumed input parameters.  \\

The values of the ANTARES environmental parameters (light absorption and scattering) 
have been measured  during several sea campaigns,  \cite{ant00, ant03, ant05}.
The uncertainty on the light absorption length in water is assumed to 
be $\pm 10\%$ over the whole wavelength spectrum and yields 
a  variation of $\pm 20\%$ on the number of expected muon events. \\

The detection efficiency  of an optical module is a function of the 
angle between its axis and the Cherenkov photon direction.
The acceptance for photons impacting the PMT at small angles with 
respect to the PMT axis (i.e. head-on) is determined with an uncertainty of $\pm15\%$. 
This is the region of the angular acceptance mostly affecting the detection of upgoing neutrino induced muons. 
At large angles (i.e. glancing), where the OMs are mostly sensitive to downgoing atmospheric muons,  
the uncertainty is about $\pm 30\%$. The global effect is a systematic uncertainty  
of $^{+35\%}_{-30\%}$  on the reconstructed track rate expectation.  
The assumed PMT effective area is that provided in the manufacturer data sheets \cite{hamamatsu}. 
It takes into account the photocathode area, the collection and the quantum efficiencies. 
A study of $^{40}$K decays observed in the ANTARES detector indicates an 
uncertainty of $\pm10\%$ on the  PMT effective area \cite{zaborov} and  
results in a variation of about $\pm20\%$ on the expected event number.
Summing  the contributions from the various
environmental and detector related systematic uncertainties 
leads to an overall uncertainty on the total number of reconstructed tracks of $^{+45\%}_{-40\%}$. \\

The reconstructed track rate is 1.52 Hz in the data. The expected track rate with the MUPAGE event generator is 1.55 Hz \cite{mupage}.
In the CORSIKA Monte Carlo samples, the reconstructed event rates are 1.1 Hz for QGSJET +  NSU  model \cite{nsu},  0.76 Hz for the QGSJET + ``poly-gonato" model  \cite{horandel}, 1.2 Hz for SIBYLL + NSU model.  There is an important dependence of the Monte Carlo prediction
on the hadronic interaction model used to describe 
the development of the air showers and on the energy spectra of primary cosmic rays. 
Within systematic uncertainties the measurements cannot 
distinguish between the  models considered.

\section{Depth-intensity relation for atmospheric muons}
The 5-line data were used to determine the depth-intensity relation of atmospheric muons, i.e. the vertical muon flux versus depth, $h$,   \cite{marco, claire}. 
Given the muon intensity, $I(\theta,{h_o})$,  at a certain depth  $h_o$ as a function of direction $\theta$,  the vertical intensity of the muon flux as a function of depth underwater  $I(\theta=0, h)$ is calculated, using the relation:
\begin{equation}
I(\theta=0, h)= I(\theta,{h_o}) \cdot |cos(\theta)| \cdot c_{corr}(\theta)
\end{equation}
where $h=h_o/cos(\theta)$ is the effective slant depth and $c_{corr}(\theta)$ is a geometrical correction factor which takes into account the curvature of the Earth, \cite{lipari, gaisser}.

The  underwater muon flux cannot be measured directly with the ANTARES detector (and in general with neutrino telescopes). It is calculated starting from the angular distribution of  reconstructed tracks, multiplied by the average multiplicity of events, estimated with Monte Carlo methods,  according to the following equation:
\begin{equation}
                       I(\theta, {h_o } ) =      { N(\theta,{h_o}) \cdot \mu(\theta,{h_o})    \over  {A_{eff }(\theta) \cdot  T \cdot \Delta \Omega(\theta)   } } 
\end{equation}
where  :
\begin{itemize}
\item $N(\theta,{h_o})$ is the number of tracks reconstructed in the angular bin around $cos(\theta)$. It is obtained from the zenith angular distribution of all reconstructed tracks and is corrected to account for the trigger and reconstruction efficiencies, calculated with Monte Carlo simulations. An iterative unfolding procedure based on Bayes theorem was applied \cite{bayes};  the resulting correction is small because of the good zenith angle resolution compared with the bin width. 
\item  $\mu(\theta,{h_o})$ is the mean muon multiplicity in the bundles, estimated from the MUPAGE simulation \cite{mupageY}. It ranges from 1.3 for vertical events to 1.1 for almost horizontal events. The CORSIKA Monte Carlo gives  $7\%$ smaller values.
\item $A_{eff }(\theta)$ is the Monte Carlo computed effective area for muons at the $\theta$ angle.
\item T and $ \Delta \Omega (\theta)$ are the live time for the selected data and the solid angle for the $ cos(\theta)$ bin, respectively. \\
\end{itemize}

Figure 4 shows the atmospheric muon flux, $I(\theta, {h_o}=2000)$, computed at 2000 m depth. It is compared with the MUPAGE simulation (solid line).

The measured muon flux, $I(\theta, {h_o } )$, was finally transformed into the vertical muon flux using eq. (1). The results are shown in Figure \ref{dir}  (black points). Statistical errors are small and not visible on the plot. The error band represents
the systematic uncertainties discussed in Section 4. 
Note that other than in the preliminary analysis described in ref. \cite{marco},
in this case no selection on the quality of the track fit is applied. 
Previous results published by ANTARES and a calculation
obtained with a parametrization taken from \cite{bugaev} are also shown. All measurements agree
within the systematic uncertainties with the model predictions.  Several
measurements from other underwater detectors are also shown for comparison. 
\begin{figure}
\begin{center}
\mbox{\includegraphics[width=11.5 cm]{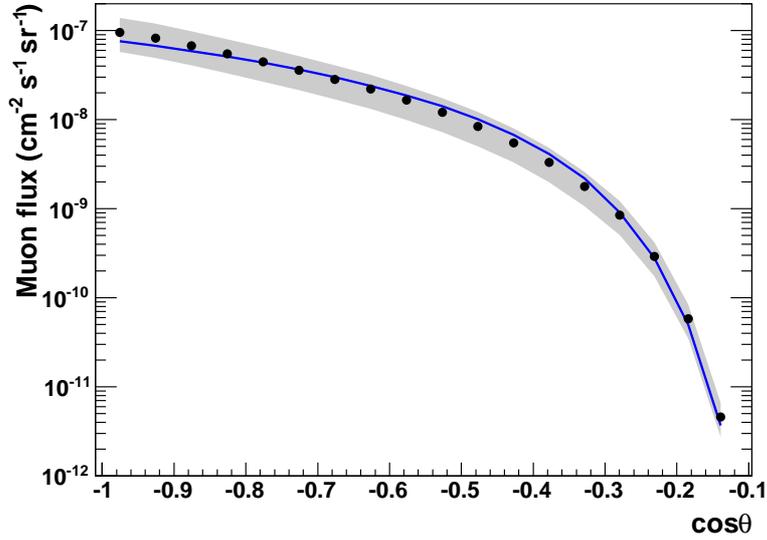}}
\caption{\label{flux}\footnotesize {The flux of atmospheric muons at 2000 m depth (black points) 
as a function of the cosine of the zenith angle. The solid line is the expectation from the MUPAGE simulation.
The grey band represents the systematic uncertainties on the measured  flux. }}
\end{center}
\end{figure}
\begin{figure}
\begin{center}
\mbox{\includegraphics[width=13.5 cm]{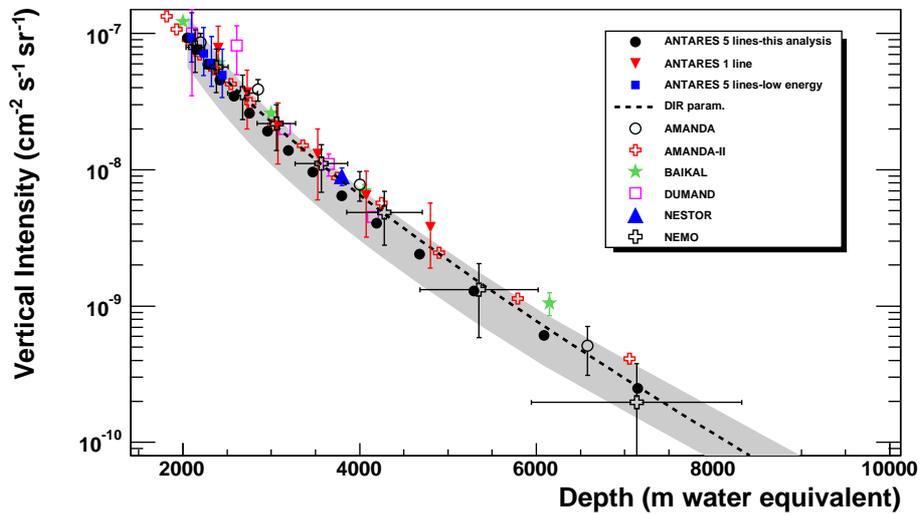}}
\caption{\label{dir}\footnotesize {
Vertical muon flux of atmospheric muons for the 5 line ANTARES data (black points) as a function of the slant depth. 
Downward triangles show the results from line 1 data \cite{line1}.  
Full squares show the results obtained with a new method and a low muon energy event selection \cite{zaborov}. 
Expectation from Bugaev parametrization (dotted line) is superimposed \cite{bugaev}. 
A compilation of results obtained with other underwater detectors is shown: AMANDA \cite{amandaB4}, AMANDA-II \cite{amandaII}   Baikal \cite{baikal}, DUMAND \cite{dumand}, NESTOR \cite{nestor}, NEMO \cite{nemo} . }}
\end{center}
\end{figure}

\section{Conclusions}
Using the data collected with the first 5 lines of the ANTARES detector, the zenith angle distribution of atmospheric muon bundles and 
the vertical muon flux as a function of the water depth were  measured. 
The results are compared with Monte Carlo expectations obtained with a 
complete Monte Carlo simulation chain  and with a parameterized 
calculation of the underwater muon flux at the detector depth. 
The major sources of systematic uncertainties in Monte Carlo simulations are the 
description of the angular dependence of the optical module efficiency and 
the theoretical models used to represent the primary cosmic ray composition
and the hadronic interactions during the air shower development.
The results of the present analysis are in agreement, 
within the systematic uncertainties, with the theoretical predictions and 
previous measurements.

\section*{Acknowledgments}
The authors acknowledge the financial support of the funding agencies:
Centre National de la Recherche Scientifique (CNRS), Commissariat
\`{a} l'Energie Atomique et aux \'{e}nergie alternatives (CEA), Agence National de la Recherche (ANR), Commission Europ\'{e}enne (FEDER fund
and Marie Curie Program), R\'{e}gion Alsace (contrat CPER), R\'{e}gion
Provence-Alpes-C\^{o}te d'Azur, D\'{e}partement du Var and Ville de La
Seyne-sur-Mer, in France; Bundesministerium f\"{u}r Bildung und
Forschung (BMBF), in Germany; Istituto Nazionale di Fisica Nucleare
(INFN), in Italy; Stichting voor Fundamenteel Onderzoek der Materie
(FOM), Nederlandse organisatie voor Wetenschappelijk Onderzoek (NWO),
in the Netherlands; 
Council of the President of the Russian Federation for young scientists and leading scientific schools supporting grants, Russia; 
National Authority for Scientific Research (ANCS) in Romania;
Ministerio de Ciencia e Innovaci\'{o}n (MICINN) and Prometeo of Generalitat Valenciana and MultiDark, in Spain.  We also
acknowledge the technical support of Ifremer, AIM and Foselev Marine
for the sea operation and the CC-IN2P3 for the computing facilities.
\bibliographystyle{model1-num-names}

\begin{thebibliography}{00}
\bibitem{amanda} M. Ackermann {\it et al.}, (AMANDA/IceCube Collab.), Astrophys. J. 675 (2008) 1014. 
\bibitem{line1} M. Ageron {\it et al.}, (ANTARES Collab.), Astropart. Phys. 31 (2009) 277.  
\bibitem{zaborov} J. A. Aguilar {\it et al.}, [ANTARES Collab.],  Astropart. Phys. 33 (2010) 86.  
\bibitem{ant02}	P. Amram {\it et al.}, (ANTARES Collab.) Nucl. Instrum. Meth. A 484 (2002) 369.  
\bibitem{ant07b}	M. Ageron {\it et al.}, (ANTARES Collab.), Nucl. Instrum. Meth. A578 (2007) 498. 
\bibitem{ant07c}	J. A. Aguilar {\it et al.}, (ANTARES Collab.), Nucl. Instrum. Meth. A570 (2007) 107. 
\bibitem{corsika} D. Heck {\it et al.}, Report FZKA 6019 (1998), Forschungszentrum Karls\-ruhe; D. Heck and J. Knapp, Report  FZKA 6097 (1998), Forschungszentrum Karls\-ruhe; http://www-ik3.fzk.de/ \~{}heck/corsika/physics\_description/corsika\_phys.html. 
\bibitem{qgsjet} N.N. Kalmykov, S.S. Ostapchenko, Yad. Fiz. 56 (1993) 105; Phys. At. Nucl. 56 N3 (1993) 346.
\bibitem{sibyll} R.S. Fletcher, T.K. Gaisser, P. Lipari, and T. Stanev, Phys. Rev. D50 (1994) 5710; J. Engel, T.K. Gaisser, P. Lipari, and T. Stanev, Phys. Rev. D46 (1992) 5013.
\bibitem{music} P. Antonioli {\it et al.},  Astropart. Phys. 7 (1997) 357. 
\bibitem{horandel} J. H\"orandel,  Astropart. Phys. 19 (2003) 193. 
\bibitem{nsu} S.I. Nikol'sky {\it et al.}, Sov. Phys. JETP 60 (1984) 10.
\bibitem{bugaev} E. V. Bugaev {\it et al.},  Phys. Rev. D58 (1998) 05401. 
\bibitem{mupage} G. Carminati {\it et al.},  Comput. Phys. Commun. 179 (2008) 915.
\bibitem{mupageY} Y. Becherini {\it et al.}, Astropart. Phys. 25 (2006) 1.
\bibitem{ant00} P. Amram {\it et al.}, (ANTARES Collab.), Astropart. Phys. 13 (2000) 127.
\bibitem{ant03}	P. Amram {\it et al.}, (ANTARES Collab.), Astropart. Phys. 19 (2003) 253. 
\bibitem{ant05}	J. A. Aguilar {\it et al.}, (ANTARES Collab.), Astropart. Phys. 23 (2005) 131. 
\bibitem{hamamatsu} http://sales.hamamatsu.com/assets/pdf/parts\_R/LARGE\_AREA\_PMT\_TPMH1286E05.pdf 
\bibitem{marco} M. Bazzotti, Ph.D Thesis, Universit$\rm{\grave{a}}$ di Bologna, 2009; M. Bazzotti for the ANTARES Collab., Proc. of the 31st ICRC, Lodz 2009,  arXiv:0911.3055 [astro-ph.HE].
\bibitem{claire} C. Picq, Ph.D Thesis, Universit$\rm{\acute{e}}$ Paris VII, 2009; Irfu report: Irfu-09-03-T. 
\bibitem{lipari} P. Lipari, Astropart. Phys. 1 (1993) 195. 
\bibitem{gaisser} T.K. Gaisser, Cosmic Rays and Particle Physics, Cambridge University Press (1990), Chapt.3. 
\bibitem{bayes} G. DÕAgostini, Nucl. Instr. Meth. A 362 (1995) 487; see alsohttp://hepunx.rl.ac.uk/adye/software/unfold/RooUnfold.html.
\bibitem{amandaB4} E. Andres {\it et al.}, (AMANDA Collab.), Astropart.Phys. 13 (2000) 1. 
\bibitem{amandaII} P. Desiati for AMANDA Collab., Proc. of the 28th ICRC, Tsukuba 2003.
\bibitem{baikal} I. A. Belolaptikov {\it et al.}, (Baikal Collab.), Astropart. Phys. 7 (1997) 263. 
\bibitem{dumand} J. Babson {\it et al.}, (DUMAND Collab.), Phys. Rev. D42 (1990) 41.
\bibitem{nestor} G. Aggouras {\it et al.}, (NESTOR Collab.), Astropart. Phys. 23 (2005) 377.
\bibitem{nemo} S. Aiello {\it et al.}, (NEMO Collab.),  Astropart. Phys. 33(2010)263.
\end{thebibliography}

\end{document}